\documentclass[conference]{IEEEtran}
\IEEEoverridecommandlockouts
\usepackage{cite}
\usepackage{amsmath,amssymb,amsfonts}
\usepackage{algorithmic}
\usepackage{graphicx}
\usepackage{textcomp}
\usepackage{xcolor}
\def\BibTeX{{\rm B\kern-.05em{\sc i\kern-.025em b}\kern-.08em
    T\kern-.1667em\lower.7ex\hbox{E}\kern-.125emX}}
\begin{document}

\def\P{ \mathbb{P} }

\def\Hmin{ H_{\min} }
\def\Cmin{ C_{\min} }
\newcommand{\krd}[1]{\textcolor{violet}{ [ #1 -- KRD ] \normalsize }}

\def\LR{\text{LR}}
\def\LLR{\text{LLR}}
\def\rLLR{L}  
\newcommand{\absLLR}[1]{|\LLR(Y_{#1})|}
\newcommand{\absorbLLR}[1]{L_{#1}}

\title{Soft detection physical layer insecurity 

}

\author{
\IEEEauthorblockN{Ken R. Duffy}
\IEEEauthorblockA{\textit{Dept. of ECE \& Dept. Mathematics} \\
\textit{Northeastern University}\\
Boston, USA \\
k.duffy@northeastern.edu}
\and
\IEEEauthorblockN{Muriel M\'edard}
\IEEEauthorblockA{\textit{Research Laboratory for Electronics} \\
\textit{Massachusetts Institute of Technology}\\
Cambridge, USA \\
medard@mit.edu}
}

\maketitle

\begin{abstract}
We establish that during the execution of any Guessing Random Additive Noise Decoding (GRAND) algorithm, an interpretable, useful measure of decoding confidence can be evaluated. This measure takes the form of a log-likelihood ratio (LLR) of the hypotheses that, should a decoding be found by a given query, the decoding is correct versus its being incorrect. That LLR can be used as soft output for a range of applications and we demonstrate its utility by showing that it can be used to confidently discard likely erroneous decodings in favor of returning more readily managed erasures. As an application, we show that feature can be used to compromise the physical layer security of short length wiretap codes by accurately and confidently revealing a proportion of a communication when code-rate is far above the Shannon capacity of the associated hard detection channel.
\end{abstract}


\section{Introduction}

Both hard- and soft-input variants of Guessing Random Additive Noise Decoding (GRAND) 
have been developed that can accurately and efficiently decode any moderate 
redundancy error correction code \cite{duffy19GRAND,duffy2021ordered,Duffy22,duffy22ORBGRAND}. GRAND algorithms operate by sequentially removing noise effect 
sequences from a hard decision sequence and querying if what remains a code-book element.
Should the sequences be ordered from most likely to least likely, based on
statistical or soft input available to the decoder, the resulting decoding is necessarily
maximum-likelihood. For reliable communication, GRAND algorithms are suitable for decoding any moderate redundancy code as 
an upper bound on their complexity can be determined as a function of the number of redundant bits added
by the code.

A natural question is how to generate soft output in the form
of a useful measure of confidence in a decoding. A simple measure is the number
of queries until a code-book element is found as it is negatively correlated to the probability
that a decoding is correct, but here we establish a more quantitative measure in the form of a log-likelihood ratio (LLR). In Forney's seminal work on list decoding \cite{Forney1968}, likelihood ratio decoding is defined in terms of the LLR of the received signal given the most-likely code-word versus the second most likely code-word. The approach is well-developed \cite{Hof2010}, and, amongst other applications, can be used to create decision regions for a list decoder to report an erasure in lieu of a likely erroneous decoding, which has applications including for hybrid automatic repeat request (ARQ). 

While the measure we introduce here to compromise physical layer security has a similar objective, it is distinct. It is the LLR between the hypothesis that a correct decoding would be identified by a given query and the hypothesis that an incorrect decoding would be identified. In contrast to the LLR considered in likelihood ratio decoding, it is calculated online during the execution of any GRAND algorithm and is informative prior to the identification of a single code-word. Moreover, when used as a measure of confidence in a decoding, it only requires the identification of one code-word rather than two or more.  

We demonstrate that the utility of the measure by illustrating its use in compromising 
physical layer security of short block-length codes. In the simple version of the wiretap channel~\cite{Wyn75, VH79}, 
Alice has data that she wishes to communicate to Bob without revealing it to Eve who has an independent, 
noisier channel than Bob's. The premise underlying operational proposals to enable physical layer security is the design of codes that are robust enough to be decodable by Bob, but whose performance degrades significantly in Eve's more noisy channel conditions \cite{TDCMM07, MV11, Demetal11}, with particular recent focus on short code-constructions \cite{ND17, nooraiepour2020secure, rana2023short}.

In the hard-detection setting, we have recently shown that unless Alice's code-rate is higher than one minus the min-entropy of the noise being experienced by Eve, in both theory and practice, Eve can use GRAND to confidently decode a proportion of Alice and Bob's communication \cite{medard22}, providing a practical mechanism to compromise physical layer security. As an illustration of the utility of the soft-output LLR measure introduced here, we demonstrate that that Eve can confidently compromise a fraction of Alice and Bob's communication far beyond her abilities in the hard detection channel. Moreover, Eve can use this attack for any short code.

\section{Guessing Random Additive Noise Decoding}
GRAND \cite{duffy19GRAND} was originally developed as a hard detection maximum likelihood (ML) decoder whose mathematical analysis provides a new approach to deriving error exponents for code-rates within capacity \cite{Gallager73} and the mirroring concept of success exponents for code-rates above capacity \cite{Ari73,dueck1979reliability}. Hardware implementations of GRAND for binary symmetric channels have been proposed and built that demonstrate efficient decoding of any moderate redundancy code \cite{abbas2020,Riaz21, Riaz22}. Algorithmic developments have further demonstrated that statistical knowledge of channel correlation statistics can be proactively exploited, obviating the need for interleaving while obtaining enhanced decoding accuracy in a fading setting \cite{An22,zhan2022noise,abbas2021high-MO}.

Soft-input variants of GRAND have been developed and analysed, ranging from those with the most significantly quantized soft input of one reliability bit per received bit \cite{Duffy22}, for which error and success exponents can be evaluated, to ones that use more detailed soft information. Ordered Reliability Bits GRAND (ORBGRAND) \cite{duffy2021ordered,duffy22ORBGRAND} is a practically realizable universal soft-input approximate ML decoder. Mathematical results prove that the basic version of ORBGRAND \cite{duffy2021ordered}, which uses a rank order of symbol reliabilities as its soft input, is almost capacity achieving in lower signal-to-noise-ratio (SNR) regimes~\cite{liu2022orbgrand}, while the most sophisticated version of ORBGRAND \cite{duffy22ORBGRAND} is almost capacity achieving for all SNR \cite{Yuan22}. In practice, ORBGRAND can provide accurate soft detection decoding of any moderate redundancy code, and published circuit designs \cite{condo2022fixed, abbas2022high} and an in silicon realization \cite{Riaz23} demonstrate that it can do so highly efficiently.

\section{GRAND decoding confidence}

For any decoding procedure, it would be desirable to have an interpretable, actionable, numerate measure of 
confidence as the decoding progresses. Such a measure could be used to adaptively inform decoding abandonment or as a soft output. Here we show that such a thing is possible for any version of GRAND, without adding computational complexity.  

The premise of the measure is to evaluate a LLR of the probability that should a decoding be found by the $q$-th query it would correct divided by the probability that it would be incorrect. To introduce the 
approach through which the LLR can be approximately calculated, we assume GRAND is querying binary sequences, although the same principle can be used for more general symbols \cite{An22,an2022soft,Chatzigeorgiou22}. 

Assume that the code-word $x^n\in\{0,1\}^n$ is transmitted and $Y^n = x^n \oplus N^n$, where $\oplus$ is addition modulo two, is received. Namely, $N^n$ is the random binary noise effect generated by the potentially continuous channel noise. For a received $Y^n$, based on channel statistics and associated soft input, if available, assume that a GRAND decoder will query the sequences $\{z^{n,1},z^{n,2},z^{n,3},\ldots\}$, $z^{n,i}\in\{0,1\}^n$, in order, where $z^{n,1}=0^n$, reflecting the fact that the hard decision sequence is the most likely one. With $G(N^n)$ denoting the number of noise effect sequences until the random noise effect $N^n$ is guessed, the probability that the true noise effect sequence would be identified within $q$ queries is
\begin{align} 
\label{eq:HP1}
\P\left(G(N^n)\leq q\right) = \sum_{j=1}^q \P(N^n=z^{n,j}).
\end{align}
For example, if the decoder's information is that bit $i$ was flipped independently with probability $B_i$, 
where soft input is typically provide in the form of the absolute value of the LLR per bit, $l_i$, so that $B_i=e^{-l_i}/(1+e^{-l_i})$,
then the likelihood of the noise effect sequence $z^n=(z_1,\ldots,z_n)$ is
\begin{align*} 
\P(N^n=z^n) &= \prod_{i=1}^n (1-B_i) \prod_{i:z_i=1} \frac{B_i}{1-B_i}
\end{align*}
and the probability accumulated after $q$ queries can be evaluated as
\begin{align} 
\label{eq:HP1b}
\sum_{j=1}^q \P(N^n=z^{n,j}) = \prod_{i=1}^n (1-B_i) \sum_{j=1}^q \prod_{i:z_i^j=1} \frac{B_i}{1-B_i}. 
\end{align}
As a result of eq. \eqref{eq:HP1b}, if bits are assumed to be flipped independently, for given $B^n$ eq. \eqref{eq:HP1} can be readily evaluated with a running sum, with one additional term per query. If no soft input is available, and instead only a statistical description is available, such as for a BSC or a channel subject to Markov bursts\cite{An22}, the evaluation $\P(N^n=z^n)$ can be based on that model.

The competing hypothesis is that an erroneous code-book element will be identified within the first $q$ queries. Let $U^n$ be the smallest number of queries in GRAND's order that would identify an erroneous decoding. While that may seem  challenging to compute $\P(U^n\leq q)$ owing to possible dependence on code-book structure, it can be universally approximated based on understanding arising from Theorem 2 of \cite{duffy19GRAND}. Essentially, with a code that takes $k$ information bits and maps them to $n>k$ coded bits, the likelihood that a random query to a random code-book would identify a code-word is $2^k/2^n$. As, by design, codewords are well-distributed within the collection of all strings of length $n$, the probability of incorrectly decoding after $q$ queries can be approximated by assuming that the code-book has been created uniformly at random and queries are made uniform at random, resulting in
\begin{align} 
\P(U^n\leq q) & \approx 1-\left(1-\frac{2^k}{2^n-1}\right)^q \nonumber\\
& \approx 1-\left(1-2^{-(n-k))}\right)^q, \label{eq:HP2}
\end{align}
which is solely a function of $n-k$ and $q$. 

\begin{figure}[htbp]
\centerline{\includegraphics[width=0.45\textwidth]{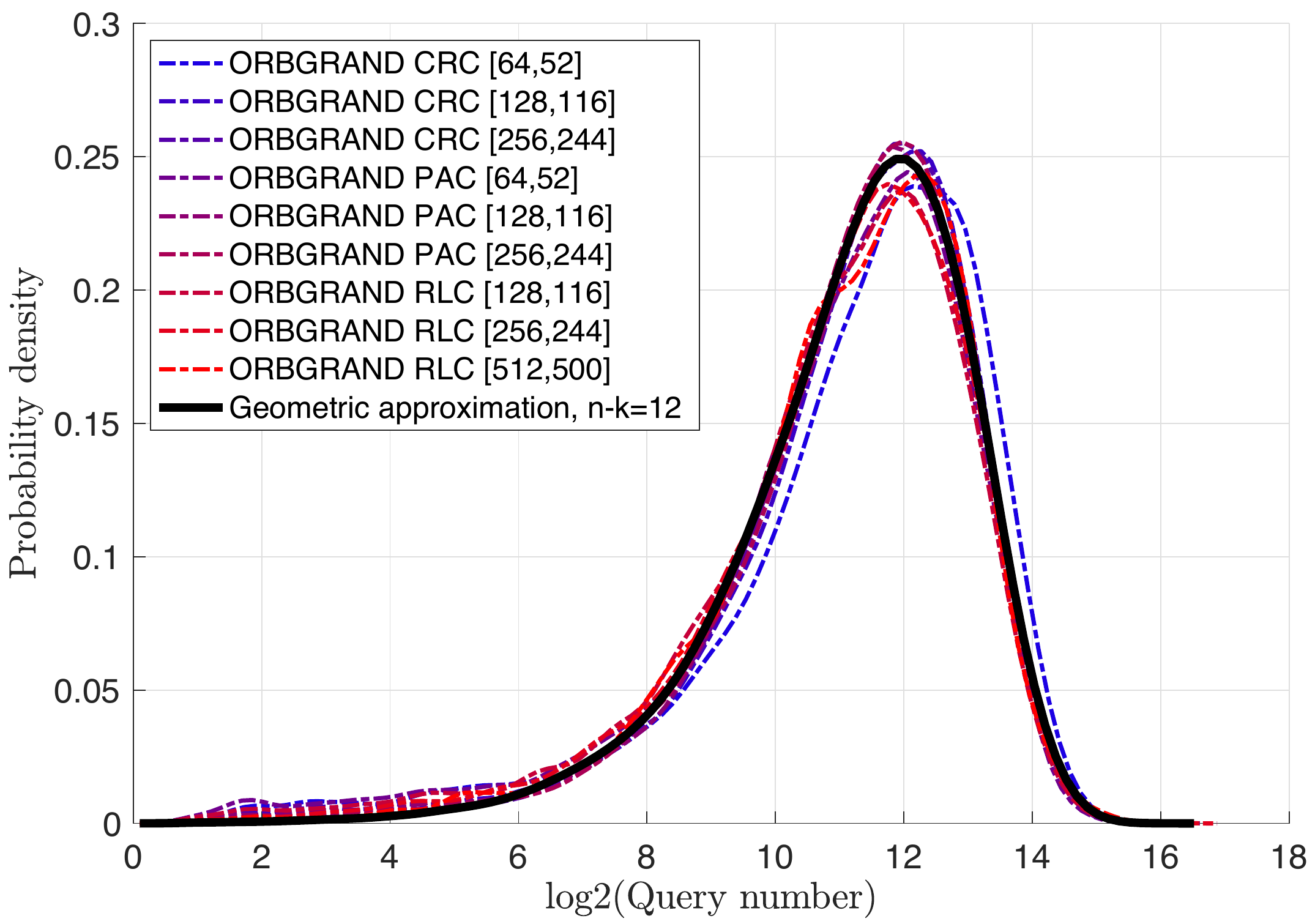}}
\caption{Empirical distribution of the number of queries until an incorrect decoding would be found by ORBGRAND for of a range of binary [n,k] codes of different length and structure, but all with $n-k=12$ in a complex additive white Gaussian (CAWGN) noise channel at an Eb/N0 of 6. In Koopman notation \cite{koopmanWeb}, the CRC polynomials are 0xbae for the 64 bit code and 0x8f3 for the 128 and 256 bit codes. The PAC codes employ the polynomial from the paper that introduced them \cite{Arikan19PAC}. The Random Linear Codes (RLCs) are created by a random binary, Bernoulli 1/2 parity matrix conditioned not to have repeated columns or all-zero rows. Also shown is geometric distribution approximation with mean $2^{n-k}$, which does not depend on code structure or channel conditions.}
\label{fig:ORB_LR_Err_12}
\end{figure}

The appropriateness and universality of the geometric distribution approximation in eq. \eqref{eq:HP2} is illustrated in Fig. \ref{fig:ORB_LR_Err_12} where it is compared with empirical results for codes of different types (Random Linear Codes, Cyclic Redundancy Check codes, and Polarization-Adjusted Convolutional codes) and different lengths (n=64, 128, 256, 512), but all with $n-k=12$. For each code, $10,000$ erroneous decodings were found with ORBGRAND, and the empirical density of $\log_2$ of the number of queries for each one is shown. Note that the approximation, and its quality, is independent of the channel conditions, and similar results are observed for different $n-k$ and additional code-structures (data not shown).

Armed with eq. \eqref{eq:HP1} and eq. \eqref{eq:HP2}, the log-likelihood ratio of the hypotheses of correct to incorrect decoding
\begin{align}
    \label{eq:LLR}
    \LLR(q) = \log_2 \frac{\P(G(N^n)\leq q)}{\P(U^n\leq q)}
\end{align}
can be approximated. As the log is base $2$, if $\LLR(q)=\tau$, then there is a $2^\tau$ to $1$ chance that if a decoding is found by the $q^\text{th}$ query, it will be correct. Not only can $\LLR(q)$ be returned when a decoding is identified, but in advance of GRAND performing the $q$-th code-book query, $\LLR(q)$ can be evaluated and a decision made as to whether to proceed with querying in the hope of identifying a code-book element, or, if the LLR is too low, abandon decoding and report an erasure as a complexity control measure rather than return an unconfident decoding. 

Note that there is no difficulty in having a mismatch between how GRAND generated the query order and the information used in the accounting in eq. \eqref{eq:HP1b}. For example, noise-effect queries could be created by ORBGRAND based on its efficient algorithm for practical convenience, but the LLR accounting could be evaluated with distinctly quantized soft information. 

\section{Compromising wiretap channels}

With all GRAND algorithms there is a negative correlation between how many queries are made until a decoding is found and the likelihood that it is correct. Consequently, they can be set to abandon decoding after a set number of queries, returning erasures rather than correct or erroneous decodings. Exploiting the mathematics of success exponents developed in \cite{duffy19GRAND}, in \cite{medard22} we demonstrate that, armed with a statistical characterisation of the channel, Eve can determine a number-of-queries threshold such that any non-abandoned decoding is likely to be correct, confidently revealing a proportion of Alice and Bob's communication for code-rates up to one minus the min-entropy of the channel noise, dubbed min-capacity. 

In the absence of soft-input, the most conservative Eve can be is to only trust the all zeros noise query corresponding to the demodulated sequence, which is a situation previously considered in the security literature \cite{smith2011}. In the presence of soft-input, we demonstrate that Eve is much more powerful as the $\LLR$ can be used to generate an abandonment criterion adapted to that particular soft input. Namely, Eve can set a threshold $\tau$ such that the decoding process is abandoned if $\LLR(q)<\tau$ as then Even estimates that there is a less than $2^\tau$ to $1$ chance that if a decoding is found by query $q$ it would be correct. As the LLR calculation is soft-input dependent, this approach inherently adapts to opportunistically allow more queries when, as informed by soft input, a noise realization happens to be good or fewer queries when a noise realization is bad.

\begin{figure*}[htbp]
\centerline{\includegraphics[width=0.9\textwidth]{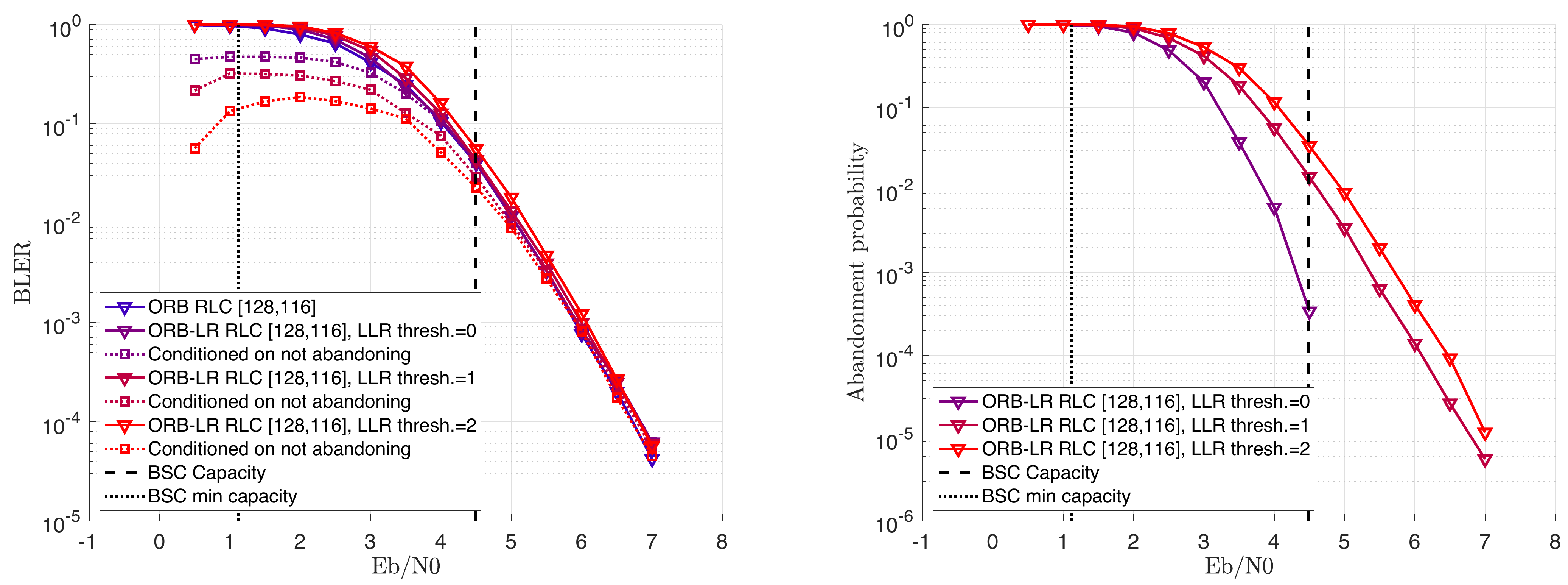}}
\caption{ORBGRAND decoding of a [128,116] RLC with LR-based abandonment threshold in a CAWGN channel with BPSK modulation. When $\LLR=\tau$, correct decoding is $2^\tau$ more likely than incorrect decoding and when $\leq \tau$, decoding is abandoned.  LHS: Block error rate as a function of Eb/N0 where dotted lines indicate the BLER conditioned on non-abandoned decodings. RHS: Proportion of abandoned decodings. Shannon capacity of the equivalent hard detection BSC is marked with a vertical dashed black line, while the dotted vertical black line indicates min-capacity, one minus the min-entropy of the equivalent BSC.
}
\label{fig:ORB_LR_RLC_128_116}
\end{figure*}

For basic ORBGRAND, which only requires the rank-ordered reliabilities of received bits as soft input and is accurate in noisy channel conditions \cite{duffy2021ordered}, but where the LLR is computed with the non-quantized soft input, the left hand side of Fig. \ref{fig:ORB_LR_RLC_128_116} provides an illustrative example of block error rate (BLER) against energy per information bit to noise power spectral density ratio (Eb/N0, in dB) in a complex AWGN channel using binary phase shift keying where abandonment thresholds that favour correct decoding by ratios of at least 1:1, 2:1 or 4:1 are employed. In the range of reliable communication, where BLER $<10^{-3}$, the performance of versions with abandonment is essentially indistinguishable from non-abandoned decoding, reflecting the fact that most noise realizations are quickly identified and a correct decoding returned. The use of an adaptive threshold has, however, converted a small proportion of erroneous decodings into abandonments, Fig. \ref{fig:ORB_LR_RLC_128_116} right hand side. 

For the corresponding hard-detection binary symmetric channel (BSC), the Eb/N0 that corresponds to where the code-rate equals the Shannon capacity is marked with a vertical dashed black line. For any Eb/N0 lower than this value, the code-rate is higher than BSC capacity. The min-capacity, one minus the min-entropy of the corresponding BSC, is marked by the vertical dotted line. In the hard-detection system, min-capacity identifies the point where the ability to correctly decode any blocks at lower Eb/N0 goes to zero if a soft-input independent query number threshold is employed \cite{duffy19GRAND,medard22}.

\begin{figure*}[htbp]
\centerline{\includegraphics[width=0.9\textwidth]{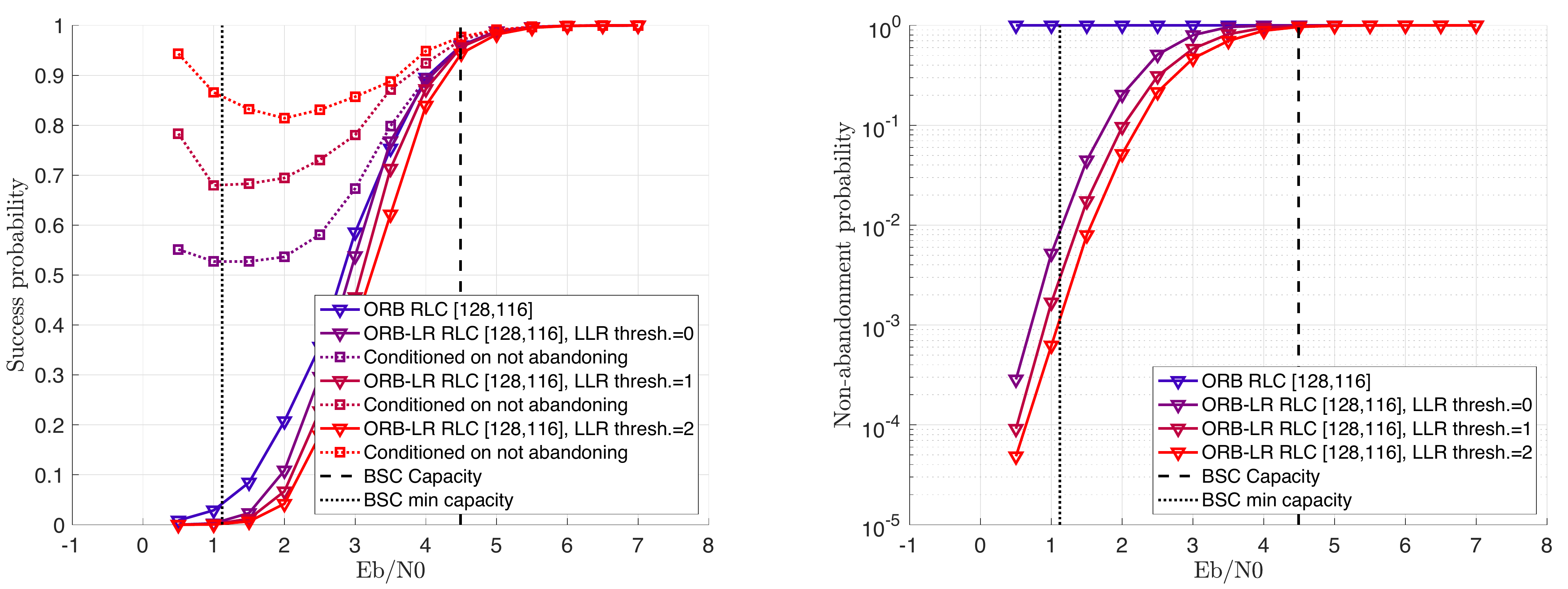}}
\caption{As in Fig. \ref{fig:ORB_LR_RLC_128_116}. LHS: shows the likelihood that the decoding is correct, 1-BLER, the success probability, where abandonment is counted as incorrect. The dashed lines are the success probability conditioned on not abandoning. RHS: the proportion of non-abandoned decodings.}
\label{fig:ORB_LR_RLC_128_116_success}
\end{figure*}

Setting an abandonment threshold based on the LLR allows Eve to change the paradigm so she is either decoding with confidence or abandoning as informed by her soft input. Fig. \ref{fig:ORB_LR_RLC_128_116_success} replots the data from Fig. \ref{fig:ORB_LR_RLC_128_116}, but focusing on what happens when the channel is noisy, i.e. Eb/N0 is below the Shannon capacity threshold of the corresponding BSC, so that most decodings of any hard detection decoder would be incorrect. By setting a $\LLR$ threshold of $\tau$, Eve estimates that a proportion of at least $2^\tau/(2^\tau+1)$ of the decodings that are not abandoned are correct. The left hand side of Fig. \ref{fig:ORB_LR_RLC_128_116_success} shows the probability of correct decoding, where abandonment counts as an error. For the LLR thresholded algorithms, the dotted lines are the proportion of decodings that are correct given non-abandonment, i.e. the proportion of correct decodings that Eve is at least $2^\tau$ to 1 confident in. By exploiting soft input, Eve can remain confident in a proportion of decodings when the code-rate is above not only Shannon capacity, but above min-capacity, corresponding to only trusting the hard decision demodulated sequence if it is in the code-book in the hard detection setting.

The right hand side of Fig. \ref{fig:ORB_LR_RLC_128_116_success} shows the proportion of decodings Eve abandons, which increases with $\tau$ and as Eb/N0 decreases. At $\sim$3dB below Shannon capacity, if Eve selects a threshold of $\tau=2$, she decodes approximately 1 in 100 packets and gets 80\% of them correct, fully compromising 1 in 125 of the packets communicated between Alice and Bob. At the min-capacity of the hard detection system, if Eve selects an LLR threshold of $0$, she decodes 1 in 100 packets and gets 50\% of them correct, compromising 1 in 200 of all blocks communicated between Alice and Bob.

\begin{figure*}[htbp]
\centerline{\includegraphics[width=0.9\textwidth]{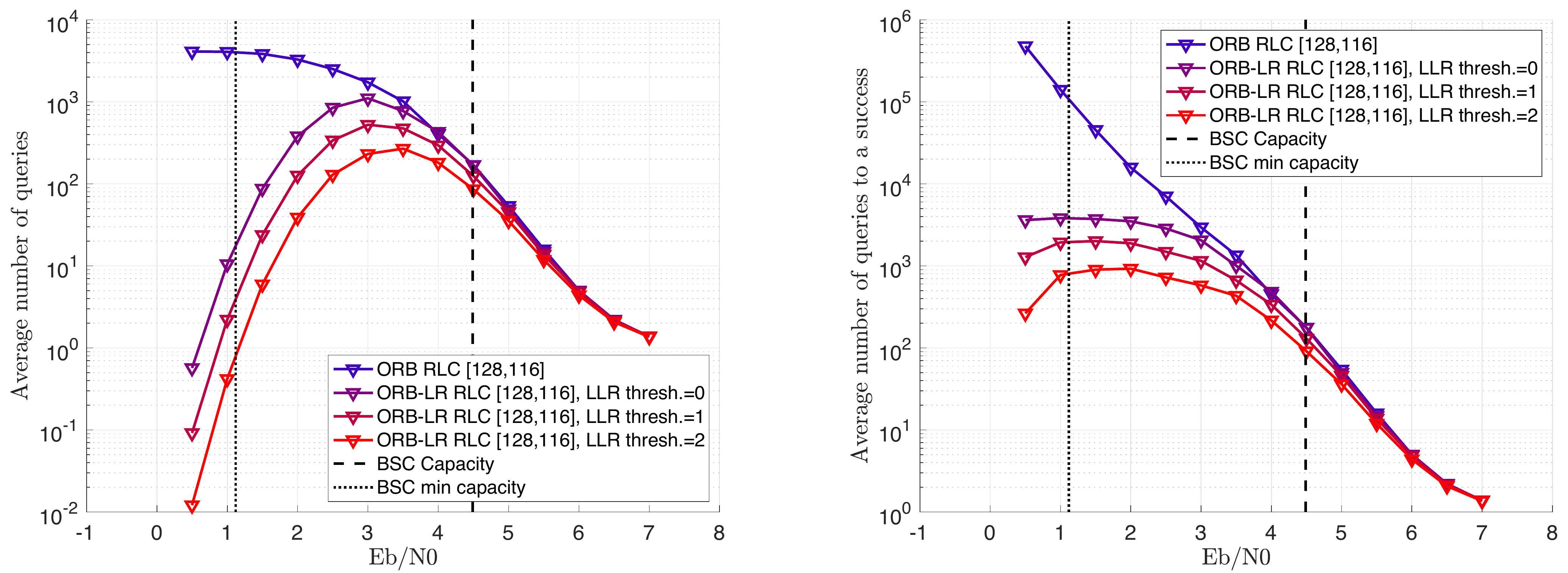}}
\caption{As in Fig. \ref{fig:ORB_LR_RLC_128_116}. LHS: average of number of queries until a decoding or abandonment. RHS: average number of queries until a successful decoding.}
\label{fig:ORB_LR_RLC_128_116_complexity}
\end{figure*}

The left hand panel of Fig. \ref{fig:ORB_LR_RLC_128_116_complexity} shows the average number of queries until a decoding is found or abandonment occurs. For ORBGRAND without abandonment, as Eb/N0 decreases to the stage where nearly all decodings are in error, the number of queries increases to the mean of the geometric distribution described in eq. \eqref{eq:HP2}, $2^{n-k} = 2^{12} \approx 10^{3.6}$. With LLR based abandonment, as the channel becomes noiser, a greater number of decoding attempts are abandoned earlier, reducing the computational burden.

The right hand panel of Fig. \ref{fig:ORB_LR_RLC_128_116_complexity} shows the average number of queries until a correct decoding is found, which accounts for the number of queries that are made leading to incorrect or abandoned decodings prior to a correct decoding being identified. By abandoning early based on an LLR threshold, Eve is doing little work in order to confidently compromise a proportion of the communication between Alice and Bob, which would be highly energy efficient in practice \cite{Riaz23}.

\begin{figure*}[htbp]
\centerline{\includegraphics[width=0.9\textwidth]{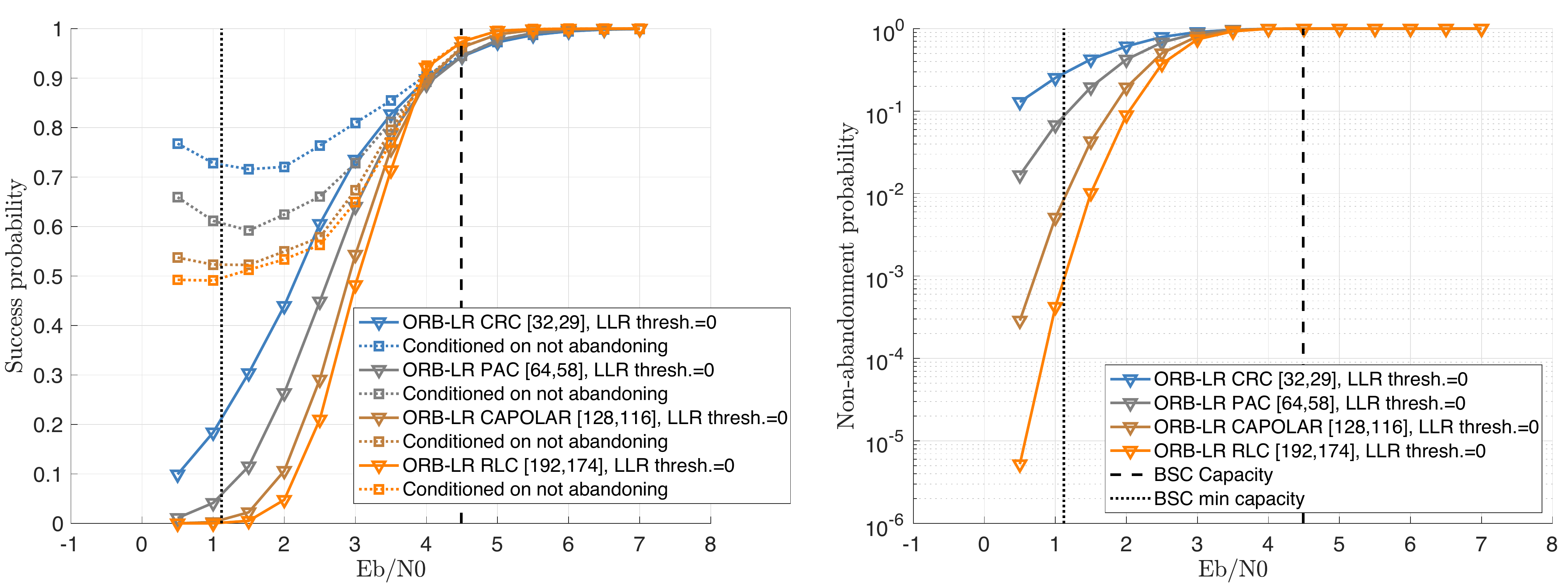}}
\caption{As in Fig. \ref{fig:ORB_LR_RLC_128_116_success} with a $\LLR$ threshold of $0$, but for codes of different structures and lengths with the same rate. CRC[32,29] code with polynomial 0x5 in Koopman notation, PAC[64,58] code with the polynomial from the original paper, CA-Polar[128,116] 5G NR code with 11-bit uplink polynomial, and a RLC[192,174].}
\label{fig:ORB_LR_X_Y_success}
\end{figure*}

Fig. \ref{fig:ORB_LR_X_Y_success} provides analogous results to those in Fig. \ref{fig:ORB_LR_RLC_128_116_success} but for four different binary linear code-structures of different dimensions, all with the same code-rate so that the hard detection Shannon capacity and min-capacity thresholds are the same. The results illustrate the fact that ORBGRAND's ability to confidently compromise a proportion of Alice and Bob's communication performance is not dependent on the code-structure independent, and more severe the shorter the code.
 
\section{Discussion}

Informative soft output from a decoding process can be used for a wide range of purposes, including in turbo-decoding updates and as confidence measures in list decoding \cite{lin2004error}. Here we have introduced a GRAND-centric soft output in the form of a LLR. While we have demonstrated its use in confidently compromising communications beyond capacity, the ability of GRAND to decode any moderate redundancy code of any length opens up other possibilities. As both hard and soft detection variants of can accurately decode CRC codes \cite{an21}, which are normally only used for error detection, this soft-output could be used to reduce the number of hybrid automatic repeat requests by not requesting retransmission when error correction with the CRC has produced a confident decoding. Moreover, as the soft output is measured in commensurate units across distinct decodings, it could also be used to identify the most confident decoding of a collection, which would be useful, for example, in selecting a lead channel in noise recycling \cite{Cohen20,riaz2022interleaved}.

The LLR in eq. \eqref{eq:LLR} is based on the cumulative likelihood that the noise-effect or erroneous decoding is found within the first $q$ queries, which could be further tailored for a list-decoding context \cite{abbas2021list}. If noise-effect sequences were produced in necessarily non-increasing order of likelihood, an alternative LLR would be to use the conditional probability that the next query would result in a correct decoding or error,
\begin{align*}
\frac{\P(G(N^n)=q)\, \P(U^n> q)}{\P(U^n=q)\, \P(G(N^n)> q)}. 
\end{align*}
That computation, however, requires comparisons of a probability mass function (PMF) for the query at which a decoding is found and with an approximate noise-effect sequence ordering, as ORBGRAND provides, some proposed noise effects can be more unlikely than those that follow later in the query order. While such queries can be rare enough not to be detrimental to decoding performance, and this measure would be more appropriate when reported as output on finding a code-word, they may be problematic when used as an abandonment condition. Within basic ORBGRAND's model, all sequences with the same logistic weight have the same likelihood, so one ameliorative approach would be to use the average likelihood of sequences with the same logistic weight in the abandonment condition. The approach taken in the results presented here is to instead use the cumulative distribution function, which  acts as an alternate form of smoothing where a single unlikely query would not result in abandonment.

\bibliographystyle{IEEEtran}
\bibliography{grand,ORBGRAND}

\end{document}